\documentclass[twocolumn,aps,prx]{revtex4-2}
\usepackage{graphicx}
\usepackage{array}
\usepackage{amsmath}
\usepackage{amsfonts}
\usepackage{hyperref}
\usepackage{xcolor}
\usepackage{physics}
\usepackage{stdclsdv}
\usepackage{amsmath, amssymb}
\hypersetup{hidelinks}

\begin{document}

\title{Generating Minimum Free Energy Paths With Denoising Diffusion Probabilistic Models}

\author{Vladimir Grigorev\\
Department of Physics and Astronomy, Johns Hopkins University, Baltimore, MD, USA}

\begin{abstract}
\noindent A method combining denoising diffusion probabilistic models (DDPMs) with the string method is presented to generate minimum free energy paths between metastable states in molecular systems. It has been demonstrated in recent work that DDPMs at low noise levels can approximate the gradient of the potential of mean force, allowing efficient sampling of high-dimensional configurational spaces. Building on this insight, it is shown here that DDPM-derived force fields accurately generate transition pathways for the analytical Müller-Brown potential and for the alanine dipeptide system at some range of noise levels for DDPMs, recovering the transition path and implicitly capturing solvent effects in the case of alanine dipeptide.

\end{abstract}

\maketitle
\section{Introduction}

\noindent Denoising diffusion probabilistic models (DDPMs) have demonstrated remarkable success in generative modeling, particularly for image synthesis \cite{ho2020denoising}. These models are part of the score-matching framework, which works by approximating gradients of log-probability distributions and bypassing direct computation of partition functions. This approach proves effective for analyzing high-dimensional probability distributions where partition function calculation is intractable, making it suitable for complex physical systems.

Although DDPMs are inspired by non-equilibrium thermodynamics \cite{sohl2015deep}, Marloes Arts et al. \cite{twoone} demonstrated that at low noise levels, the predicted score function of DDPM can approximate a (mean) force field, enabling direct simulation of coarse-grained molecular dynamics. Building on this insight, we integrate DDPMs with the string method \cite{weinan2002string} to generate minimum free energy paths (MFEPs) between metastable states, where free energy means the potential of the mean force $W(x)$. These paths characterize the maximally probable routes connecting stable configurations, providing crucial insights into transition mechanisms.

We applied this methodology for the analytical Müller-Brown potential and the alanine dipeptide system, demonstrating that DDPM-derived force fields can accurately identify minimum free energy paths in systems of varying complexity.

\begin{figure}[t]%
\centering
\includegraphics[width=0.45\textwidth]{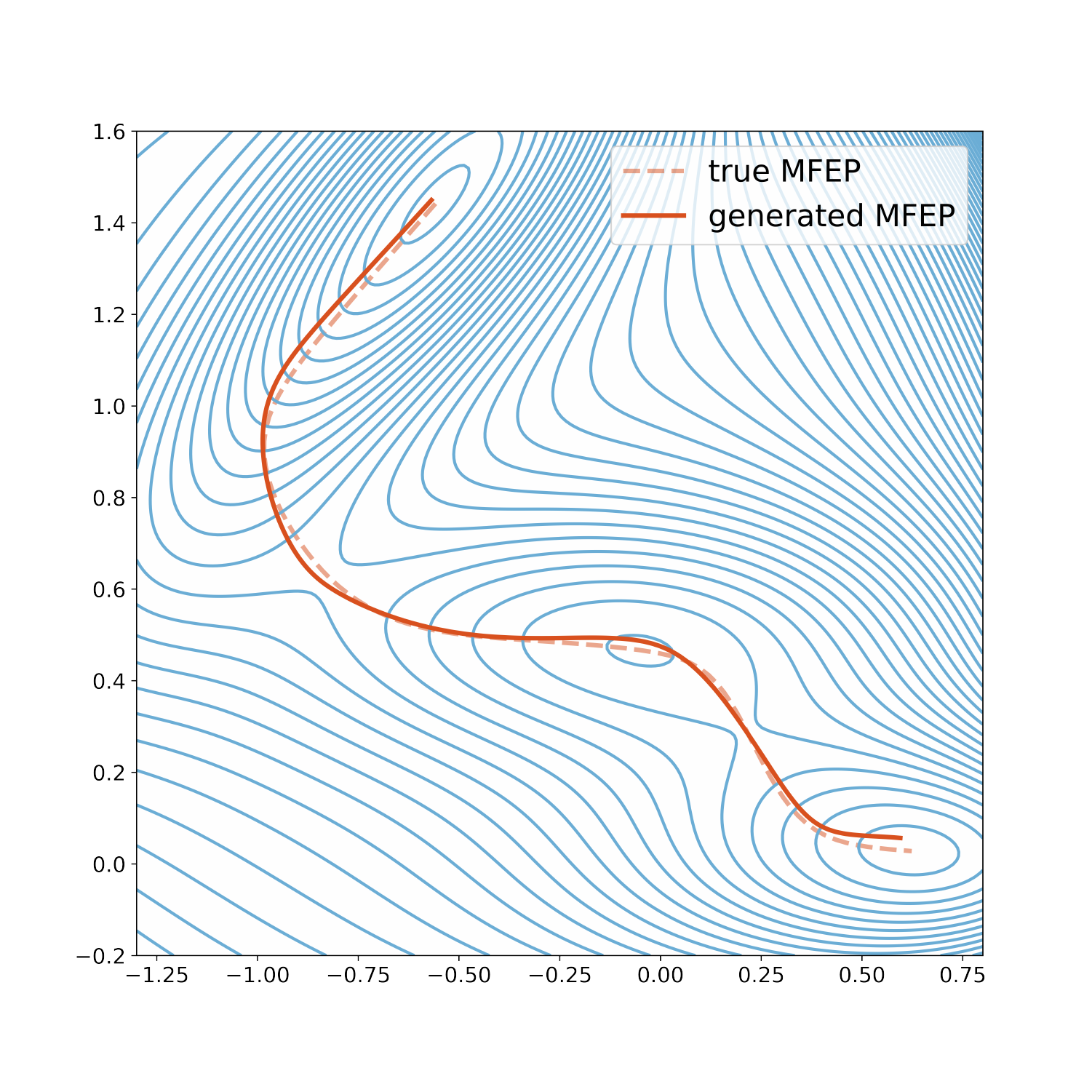}
\caption{Example of a generated MFEP for the analytical Müller-Brown potential at noise level $t = 29$ (See Experiments for details).}\label{fig1}
\end{figure}

\section{Background}

Diffusion models \cite{sohl2015deep} consist of two main processes, diffusion and denoising. For the diffusion process, starting from the sample $\mathbf{x}_0$ from the unknown distribution $q(\mathbf{x}_0)$, we gradually add noise to it in the form of a Markov chain according to a variance schedule $\beta_1, \dots, \beta_T$:

\[
\begin{array}{cc}
& q(\mathbf{x}_{1:T}|\mathbf{x}_0) = \prod_{t=1}^{T} 
\displaystyle q(\mathbf{x}_t|\mathbf{x}_{t-1}), \\[10pt]
& q(\mathbf{x}_t|\mathbf{x}_{t-1}) = \mathcal{N}(\mathbf{x}_t; \sqrt{1 - \beta_t} \, \mathbf{x}_{t-1}, \beta_t \mathbf{I}) ,
\end{array}
\]
\\

The denoising process distributions are chosen to have the same functional form with learned Gaussian transitions starting at $p(\mathbf{x}_T) = \mathcal{N}(\mathbf{x}_T; \mathbf{0}, \mathbf{I})$:

\[
\begin{array}{cc}
& p_\theta(\mathbf{x}_{0:T}) = p(\mathbf{x}_T) \prod_{t=1}^{T} p_\theta(\mathbf{x}_{t-1}|\mathbf{x}_t), \\[10pt] &
 p_\theta(\mathbf{x}_{t-1}|\mathbf{x}_t) = \mathcal{N}(\mathbf{x}_{t-1}; \boldsymbol{\mu}_\theta(\mathbf{x}_t, t), \boldsymbol{\Sigma}_\theta(\mathbf{x}_t, t)),
\end{array}
\]
\\

The following parametrization suggested by Ho et al. \cite{ho2020denoising} significantly enhances the performance:

\[
\begin{array}{cc}
     & \boldsymbol{\Sigma}_\theta(\mathbf{x}_t, t) = \beta_t \mathbf{I}, \\[10pt]
     & \boldsymbol{\mu}_{\theta}(\mathbf{x}_{t}, t) = \frac{1}{\sqrt{\alpha_{t}}} \left( \mathbf{x}_{t} - \frac{\beta_{t}}{\sqrt{1 - \bar{\alpha}}_{t}} \boldsymbol{\epsilon}_{\theta} (\mathbf{x}_{t}, t) \right),
\end{array}
\]

Here $\boldsymbol{\epsilon}_{\theta}$ is the output of the neural network, learned by minimizing the loss.

\[
{L}=\mathbb{E}_{t, \mathbf{x}_{0},\mathcal{N}(\boldsymbol{\epsilon}; \mathbf{0}, \mathbf{I})}\left[ \Vert \boldsymbol{\epsilon} - \boldsymbol{\epsilon}_{\theta}(\sqrt{\bar{\alpha}_{t}}\mathbf{x}_{0} + \sqrt{1 - \bar{\alpha}_{t}}\boldsymbol{\epsilon}, t) \Vert^{2} \right].
\]

Marloes Arts et al. \cite{twoone} showed that at sufficiently low noise levels,
\begin{equation*}
\nabla_\mathbf{x}\log q(\mathbf{x}) = -\frac{1}{\sqrt{1-\bar{\alpha}_{t}}}\boldsymbol{\epsilon}_{\theta}(\mathbf{x},t),
\end{equation*}
and that with prior training of the neural network $\boldsymbol{\epsilon}_{\theta}$, the system's dynamics can be reproduced via Langevin (or Brownian) dynamics.

For molecular systems, it is advantageous to train the neural network on the directed relative distances of a subset of relevant atoms. This approach, while respecting molecular symmetries, constitutes a form of dimensionality reduction that preserves the dynamics of the selected subset as if it was modeled within the full system but at an exponentially lower computational cost. This reduction in computational demand is, in fact, the primary motivation for employing DDPMs to reproduce the dynamics of molecular systems.

Metastable states of a system are found around the minima of the potential of mean force \( W(\mathbf{x}) \). The minimum free energy path that connects these states is, by definition, a smooth curve \( \gamma \) that satisfies
\begin{equation*}
    \left( \nabla W \right)^\perp (\gamma) = 0,
\end{equation*}
where \( \left( \nabla W \right)^\perp \) is the component of \( \nabla W \) orthogonal to \( \gamma \).

To locate MFEP between metastable states in the DDPM-derived potential landscape, we implement the string method \cite{weinan2002string}. The continuous path $\gamma$ is discretized into a finite set of images $\mathbf{x}_1, \dots, \mathbf{x}_j, \dots, \mathbf{x}_J$ and evolves them according to the external mean force field. The discretized string evolves under the influence of the underlying free energy landscape, which can be approximated with a DDPM-derived force field (in units of $k_{\text{B}} \mathcal{T}$ where $\mathcal{T}$ is temperature, which is absorbed into the time step $\Delta \tau$ for the string method evolution):
\begin{equation*}
-\nabla W_t(\mathbf{x}_j) = - \frac{1}{\sqrt{1-\bar{\alpha}_t}}\boldsymbol{\epsilon}_\theta(\mathbf{x}_j,t)
\end{equation*}
The timestep $t$ (the noise level) in the diffusion process determines the level of noise added to the system, where $t=T$ corresponds to maximum noise (Gaussian sample) and $t=1$ to minimum noise. Theoretically, the lowest noise level should give the most accurate representation of the original distribution. However, empirically, intermediate time steps have been shown to often provide better estimates of the equilibrium distribution \cite{twoone}. In the context of generating MFEPs, using intermediate timesteps can also be beneficial because the string method converges faster on smoother potential surfaces, though this comes at the cost of reduced accuracy in representing the fine details of the free energy landscape. 

We evolve the string according to the forward Euler method:
\begin{align*}
\mathbf{\Tilde{x}}_j^{i+1}  = \mathbf{x}_j^{i} -\nabla W_t(\mathbf{x}_j^i) \Delta \tau
\end{align*}
After each step $i$, reparametrization of the string is required to maintain equal spacing between images through linear interpolation, for example. We first construct a piecewise linear interpolation for given images $\mathbf{\Tilde{x}}_1^{i+1}, \dots, \mathbf{\Tilde{x}}_j^{i+1}, \dots, \mathbf{\Tilde{x}}_J^{i+1}$ and reparametrize the string by sampling equidistant points from it. The resulting discretized path $\mathbf{x}_1^{i+1}, \dots, \mathbf{x}_j^{i+1}, \dots, \mathbf{x}_J^{i+1}$ preserves the geometry of the original path while ensuring uniform spacing between consecutive points. The algorithm continues until convergence, defined by the maximum displacement of any image falling below a specified tolerance:
\begin{equation*}
\max_{j} \|\mathbf{x}_j^{i+1} - \mathbf{x}_j^i\| < \text{tolerance}
\end{equation*}

\section{Experiments}

\subsection{Analytical potential} 

\begin{figure}[t]%
\centering
\includegraphics[width=0.35\textwidth]{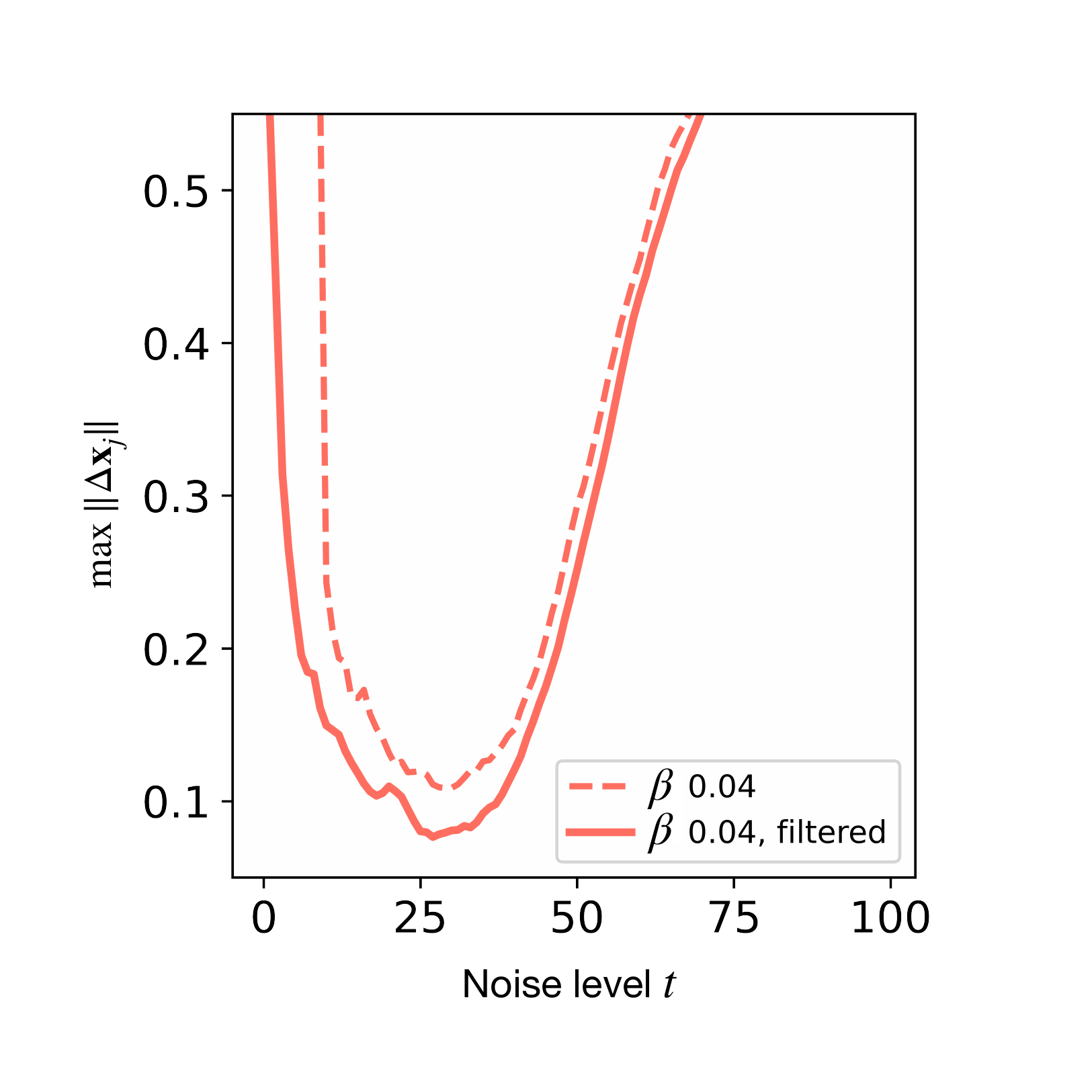}
\caption{Accuracy of generated paths for various noise levels $t$, measured by the maximum displacement between converged string images and corresponding true string images. The dashed line corresponds to an unmanipulated generated dataset. The solid line corresponds to free energy-filtered datasets containing full MFEP. Both curves average predictions from 10 independently trained neural networks.}\label{fig2}
\end{figure}

The first example utilizes the Müller-Brown potential, which serves as a free energy landscape in this case. The dataset was generated using the Metropolis algorithm, which targeted the Boltzmann distribution, where the acceptance probability is given by equation \( P(\text{accept}) = \min(1,e^{-\beta \Delta W} )\). In this equation, \( \beta \) represents the inverse temperature, and \( \Delta W \) is the difference in free energy between states. A dataset comprising \( 5 \times 10^5 \) points was sampled to ensure sufficient coverage of the free energy surface.

Following data collection, a multilayer perceptron was implemented with $t$-dependent scaling through trainable embeddings, each initialized from a uniform distribution. The training incorporated a progressive denoising strategy, where the maximal noise level of $T = 100$.

As a metric for describing MFEP accuracy, we used the maximal displacement between the converged string images and the corresponding true string images obtained via the string method using actual log-probability gradients. Experiments with the analytical potential indicate an empirically derived optimal noise level around $t = 30$, where the path deviation is minimized.

The energy filtering procedure, which consists of only keeping samples with free energy below a certain threshold ($-30$ in this specific case), gives slightly better results on the chosen metric. This can be explained by the neural network focusing more attention on the region of interest, where the free energy is lower, which likely contains the MFEP. However, this threshold should be monitored carefully, as setting it too high might result in missing parts of the MFEP.

\subsection{Alanine dipeptide} 
Similarly, the same procedure can be applied to a molecular system, such as solvated alanine dipeptide. The dataset was generated using molecular dynamics simulations performed in GROMACS \cite{van2005gromacs} with Amber03 force field. The molecule was placed in a cubic box with 1.2 nm padding and solvated with explicit water molecules with a NaCl concentration of 0.15 M. The system underwent energy minimization followed by NVT and NPT ensemble equilibration. Simulations were performed using the leap-frog integrator for 100 ns with a 2 fs timestep. Temperature coupling was achieved using a velocity-rescale thermostat (temperature = 300K, relaxation time = 0.1 ps). The pressure was maintained at 1 bar using the Parrinello-Rahman barostat (relaxation time = 2.0 ps). Trajectory data was saved every 2 ps.

\begin{figure}[t]%
\centering
\includegraphics[width=0.45\textwidth]{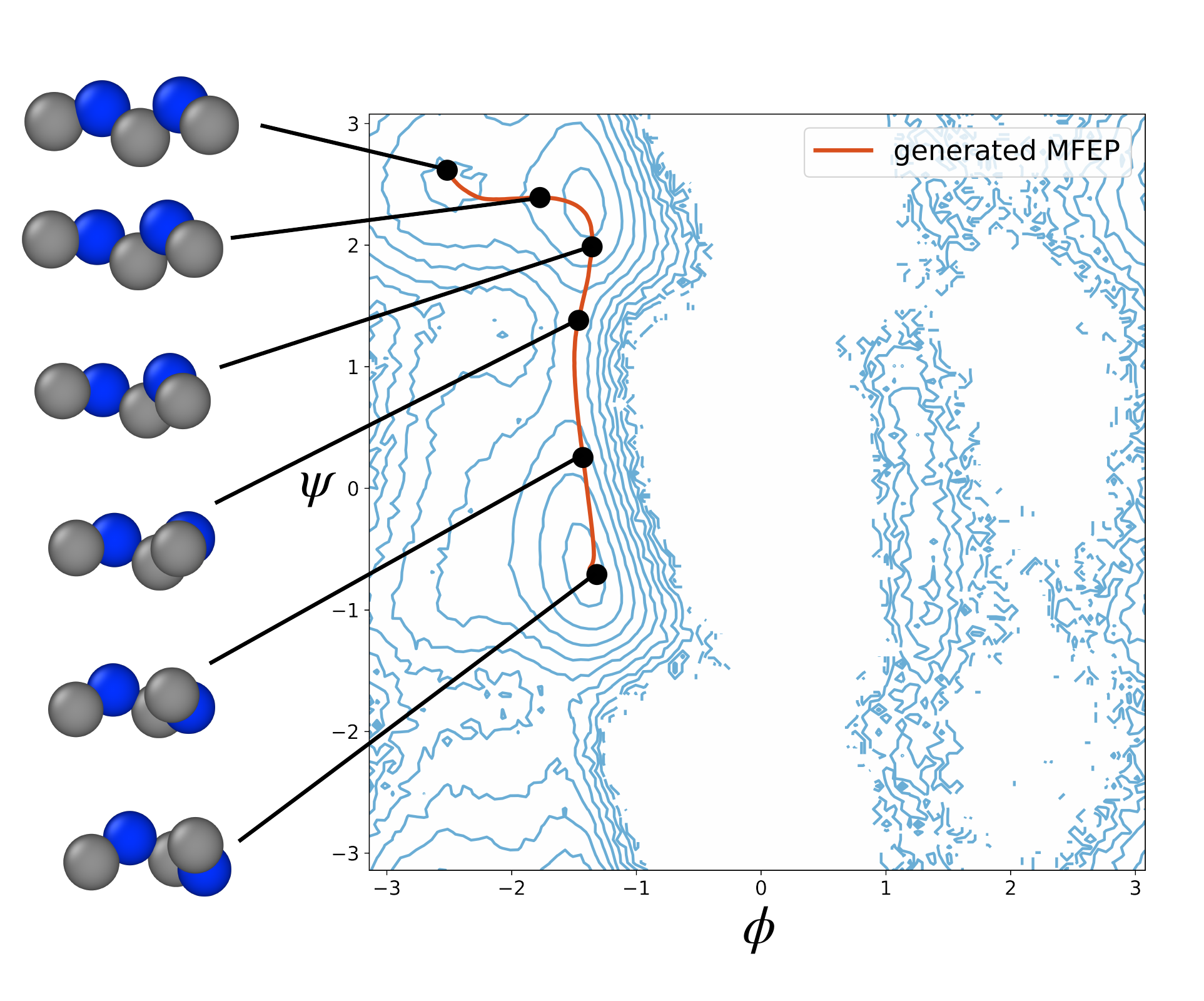}
\caption{Generated MFEP connecting two metastable states $C_5$ and $\alpha_R$ of alanine dipeptide at noise level $t = 50$ displayed on the Ramachandran plot.}\label{fig3}
\end{figure}

Following the simulation, only the coordinates of the five central backbone atoms were kept, representing substantial dimensionality reduction from a solvated all-atom system (thousands of degrees of freedom to fifteen). The dataset was filtered based on the probability density of the Ramachandran plot, retaining only configurations with probability densities higher than the average. 

The noise prediction network utilized a multilayer perceptron architecture. The input consisted of directed edges between five selected backbone atoms with a learned 64-dimensional $t$ embedding, and the output was the predicted noise for each atomic position. The network was trained on the filtered dataset according to the procedure outlined in the background section with a maximal noise level $T = 1000$ and then used to generate MFEP connecting the $C_5$ and $\alpha_R$ states of alanine dipeptide. Initial coordinates for these states were selected from the dataset and aligned through rotation to minimize mean squared displacement along the path, this removes unnecessary rotation in the generated MFEP.

The resulting MFEP covered a metastable water-stabilized \cite{rubio2017effect} polyproline II (PII) conformation, demonstrating the implicit learning of solvation effects.

\section*{Code Availability}
The implementation of the methods described in this paper is available on the \href{https://github.com/Grivov/Generating-Minimum-Free-Energy-Paths-With-Denoising-Diffusion-Probabilistic-Models}{GitHub repository}.

\bibliographystyle{apsrev4-2}

\end{document}